\documentclass[preprint,12pt]{elsarticle}
\usepackage{amssymb}

\journal{J. Alloys Compd.}

\begin{document}

\begin{frontmatter}

\title{Superconductivity and hardness of the equiatomic high-entropy alloy HfMoNbTiZr}

\author{Jiro Kitagawa$^1$}

\address{$^1$ Department of Electrical Engineering, Faculty of Engineering, Fukuoka Institute of Technology,3-30-1 Wajiro-higashi, Higashi-ku, Fukuoka 811-0295, Japan}
\ead{j-kitagawa@fit.ac.jp}

\author{Kazuhisa Hoshi$^2$}

\address{$^2$ Department of Physics, Tokyo Metropolitan University,Hachioji 192-0397, Japan}

\author{Yuta Kawasaki$^3$}

\address{$^3$ Department of Electrical Engineering, Faculty of Science and Engineering, Kyushu Sangyo University, 2-3-1 Matsukadai, Higashi-ku, Fukuoka 813-8503,Japan}

\author{Rikuo Koga$^1$}

\address{$^1$ Department of Electrical Engineering, Faculty of Engineering, Fukuoka Institute of Technology,3-30-1 Wajiro-higashi, Higashi-ku, Fukuoka 811-0295, Japan}

\author{Yoshikazu Mizuguchi$^2$}

\address{$^2$ Department of Physics, Tokyo Metropolitan University,Hachioji 192-0397, Japan}

\author{Terukazu Nishizaki$^3$}

\address{$^3$ Department of Electrical Engineering, Faculty of Science and Engineering, Kyushu Sangyo University, 2-3-1 Matsukadai, Higashi-ku, Fukuoka 813-8503,Japan}

\begin{abstract}
We have found that body-centered cubic (bcc) HfMoNbTiZr is a type-II BCS
high-entropy alloy (HEA) superconductor with a superconducting critical temperature of $T_\mathrm{c}$=4.1 K. By employing a Debye temperature $\theta_\mathrm{D}$ of 263 K and $T_\mathrm{c}$, the electron-phonon coupling constant $\lambda_\mathrm{e-p}$ is calculated to be 0.63.
The electronic structure calculation revealed band broadening with energy uncertainties due to atomic disorders.
The superconducting properties are compared among equiatomic quinary bcc HEA superconductors.
$T_\mathrm{c}$ decreases with decreasing $\lambda_\mathrm{e-p}$, which is negatively correlated with $\theta_{D}$.
The negative correlation between $\lambda_\mathrm{e-p}$ and $\theta_{D}$ would be caused by a shorter phonon lifetime at a higher $\theta_{D}$, which is based on phonon broadening due to atomic disorder and the uncertainty principle.
The Vickers microhardness was measured for several bcc HEA superconductors.
$T_\mathrm{c}$ is exceptionally low in a superconductor with relatively high hardness, because the hardness generally increases with increasing $\theta_\mathrm{D}$, and $\theta_\mathrm{D}$ is negatively correlated with $T_\mathrm{c}$ in the high-entropy state.
\end{abstract}




\end{frontmatter}


\section{Introduction}
High entropy alloys (HEAs) have attracted much attention due to their unconventional alloy design principles and superior physical properties\cite{Wang:JMCA2021,Sathiyamoorthi:PMS2022,Chaudhary:MT2021}.
In HEA, more than five principal elements form a solid solution, in which the configurational entropy due to atomic disorders increases.
HEAs show an increase in the solid solution phase forming ability, which is called the high-entropy effect and leads to outstanding thermal stability.
The high-entropy concept has been extended to various materials such as oxides, chalcogenides, borides, carbides, and nitrides \cite{Musico:APLMater2020,Ying:JACS,Iwan:AIPAdv2021,Wang:AAC2022,Lu:ASS2021}.
Due to good mechanical properties, such as high fracture toughness and high yield strength, HEAs are considered promising materials for structural use\cite{Li:PMS2021}.
Moreover, HEAs show rich functionality, such as energy storage, radiation protection, magnetic refrigeration, superconductivity, and biocompatibility\cite{Wang:JMCA2021,Marques:EES2021,Pickering:Entropy2021,Castro:Metals2021}.
The cocktail effect, which means an enhancement of physical properties beyond the simple mixture of those of constituent elements, is one of the central issues of HEAs.
This effect is found in HEAs that show catalytic\cite{Wu:JACS2020}, thermoelectric\cite{Jiang:Science2021}, magnetic\cite{Marik:ScrMater2020}, and capacitive energy storage\cite{Yang:NM2022} behaviors.
The other current issue is the advancement of functionality based on the large compositional space, which is the characteristic feature of HEAs.
HEAs can easily exhibit a microstructure in a large compositional space.
Controlling
the microstructure improves the mechanical or magnetic properties\cite{Bhardwaj:TI2021,Rao:AFM2021}.
Materials research on HEAs showing novel phenomena\cite{Zherebtsov:Int,Kitagawa:APLMater2022} by utilizing the large compositional space is also attractive and carried out worldwide.

Materials research on HEA superconductors has been a growing research area since the discovery of the body-centered cubic (bcc) HEA superconductor Ta$_{34}$Nb$_{33}$Hf$_{8}$Zr$_{14}$Ti$_{11}$ in 2014\cite{Kozelj:PRL2014,Sun:PRM2019,Kitagawa:Metals2020}.
HEA superconductivity is now found in various crystal structures: bcc\cite{Rohr:PRM2018,Marik:JALCOM2018,Ishizu:RINP,Harayama:JSNM2021,Sarkar:IM2022,Motla:PRB2022}, hexagonal close-packed (hcp)\cite{Lee:PhysicaC2019,Marik:PRM2019}, face-centered cubic (fcc)\cite{Zhu:JALCOM:2022}, CsCl-type\cite{Stolze:ChemMater2018}, A15\cite{Wu:SCM2020,Yamashita:JALCOM2021}, NaCl-type\cite{Mizuguchi:JPSJ2019,Yamashita:DalTran2020}, $\alpha$-Mn-type\cite{Stolze:JMCC2018}, $\sigma$-phase\cite{Liu:ACS2020}, CuAl$_{2}$-type\cite{Kasen:SST2021}, BiS$_{2}$-based, and YBCO-based\cite{Sogabe:SSC2019,Shukunami:PhysicaC2020} structures.
Several interesting superconducting properties have been reported, such as
the robustness of superconductivity against extremely high pressure\cite{Guo:PNAC2017} or magnetism\cite{Liu:JALCOM2021} and the cocktail effect in enhancing bulk superconductivity\cite{Sogabe:SSC2019}.
The fabrication of HEA thin-film superconductors is one of the frontier studies\cite{Zhang:PRR2020,Shu:APL2022}.
It is also attractive that HEA superconductors display a high critical current density by improving the thermal treatment\cite{Gao:APL2022} or sample preparation method\cite{Kim:ActMater2022,Jung:NC2022} when viewed from their applicability.

In bcc HEA superconductors, the high-entropy effect on superconducting properties has been investigated, but a clear conclusion has not been obtained\cite{Sun:PRM2019,Kitagawa:Metals2020,Rohr:PRM2018,Rohr:PNAS2016}.
Previous papers have focused on the configurational entropy dependence of superconducting properties\cite{Sun:PRM2019,Kitagawa:Metals2020,Rohr:PRM2018,Rohr:PNAS2016}.
The configurational entropy $\Delta S_\mathrm{mix}$ is expressed by the equation:
\begin{equation}
\Delta S_\mathrm{mix}=-R\sum_{i=1}^{n}c_{i}\mathrm{ln}c_{i}
\end{equation}
where $n$ is the number of elements, $c_{i}$ is the atomic fraction, and $R$ is the gas constant.
In the quinary alloy, $\Delta S_\mathrm{mix}$ reaches the maximum value of 1.61$R$ when the equiatomic state is achieved.
We have changed the strategy of investigating the high-entropy effect on superconducting properties and attempted to discuss the material dependence of superconductivity among equiatomic quinary alloys with the maximized $\Delta S_\mathrm{mix}$.
However, equiatomic quinary bcc HEA superconductors are rare and have been reported only in HfNbTaTiZr\cite{Vrtnik:JALCOM2017}, HfNbReTiZr\cite{Marik:JALCOM2018}, and HfNbTaTiV\cite{Sarkar:IM2022}.
HfMoNbTiZr is also an equiatomic bcc HEA\cite{Guo:MD2015,Tseng:Entropy2019}, but its low-temperature physical properties have not been investigated.
We have found superconductivity in HfMoNbTiZr.
Therefore, the first purpose of this study is to characterize the fundamental superconducting properties of HfMoNbTiZr and discuss the material dependence of superconductivity among equiatomic quinary alloys.

The superconductivity combined with high hardness was recently studied in transition metal carbide-based materials\cite{Ge:ICF2019,Cui:JALCOM2021}.
Such a material would be applied as a multifunctional superconductor.
Considering that many HEAs have the property of high hardness, the investigation of the relationship between the hardness and the superconductivity in bcc HEA superconductors would be meaningful in designing a multifunctional HEA superconductor.
In addition, the hardness is generally related to the phonon information.
Because bcc HEA superconductors follow the BCS theory based on electron-phonon coupling, the hardness study is also important to obtain deeper insight into the superconductivity in a high-entropy state.
Therefore, the second purpose of this study is the investigation of the relationship between the hardness and the superconductivity in bcc HEA superconductors, including HfMoNbTiZr and nonequiatomic superconductors reported by our team\cite{Ishizu:RINP,Harayama:JSNM2021}.

In this paper, we investigate the fundamental superconducting properties of HfMoNbTiZr with a superconducting critical temperature $T_\mathrm{c}$ of 4.1 K by measuring magnetization, electrical resistivity, and specific heat.
The electronic structure calculation was also conducted to elucidate the contribution of each element to the density of states at the Fermi level and the effect of atomic disorders.
The comparison of superconducting properties among equiatomic quinary bcc HEA superconductors was discussed.
The Vickers microhardness was measured for several bcc HEA superconductors.
The valence electron count per atom (VEC) dependence of hardness and the effect of hardness on the superconductivity were discussed.

\section{Materials and Methods}
Polycrystalline sample was prepared by a homemade arc furnace using the constituent elements Hf (98 \%), Mo (99.9 \%), Nb (99.9 \%), Ti (99.9 \%), and Zr (99.5 \%) under an Ar atmosphere.
The sample was flipped and remelted several times to ensure homogeneity.
The button-shaped sample with a mass of 2 g was quenched on a water-chilled Cu hearth.
In this study, the sample received neither heat treatment nor deformation (e.g., rolling).

Room-temperature X-ray diffraction (XRD) patterns were obtained using an X-ray diffractometer (XRD-7000L, Shimadzu) with Cu-K$\alpha$ radiation in a Bragg-Brentano geometry.
We used thin slabs cut from the sample due to its high hardness.
Scanning electron microscopy (SEM) images were collected by field-emission scanning electron microscopy (FE-SEM; JSM-7100F, JEOL).
The chemical composition in each sample area was evaluated by an energy-dispersive X-ray (EDX) spectrometer equipped with FE-SEM by averaging several data collection points.
The elemental mappings were also obtained by the EDX spectrometer.

The temperature dependence of dc magnetization $M$($T$) and the isothermal magnetization curve were measured by a SQUID magnetometer (MPMS3, Quantum Design).
A sample with a weight of 9.2 mg was used in these measurements.
A parallelepiped sample with dimensions of 1.47$\times$1.02$\times$8.32 mm$^{3}$ was used for the measurement of electrical resistivity $\rho$ carried out by a PPMS (Quantum Design).
The distance between the voltage electrodes was 3.91 mm.
The specific heat was measured by the thermal-relaxation method using the PPMS apparatus at 0 T and 9 T.
The sample mass was 15.894 mg.
The Vickers microhardness was measured on polished cross-sectional surfaces under a 300 g load applied for 10 s using a Shimadzu HMV-2T microhardness tester.
All samples used in the measurements of physical properties lay near one another in the button-shaped sample.

To calculate the electronic structure of materials with atomic disorder, the coherent potential approximation (CPA) approach is efficient without the necessity of a large supercell.
In this study, we used the Akai-KKR program package\cite{Akai:JPSJ1982,Fukushima:PRM2022}, which employs the Korringa-Kohn-Rostoker (KKR) method
with CPA.
We adapted the generalized gradient approximation from Perdew-Burke-Ernzerhof (PBE).
The spin-orbit interaction was considered.
\section{Results and Discussion}
\subsection{Structural and Metallographic Characterizations}
\begin{figure}
\begin{center}
\includegraphics[width=1\linewidth]{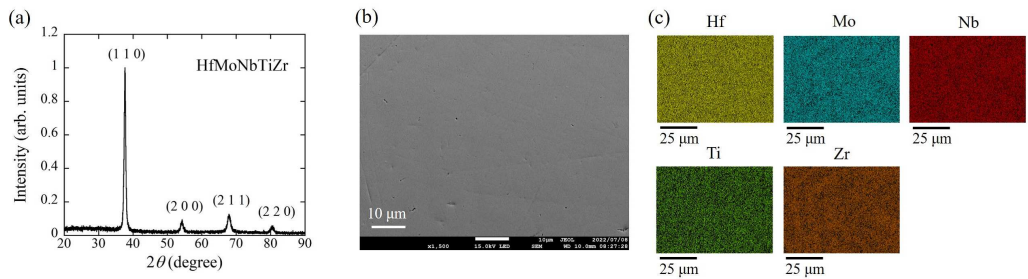}
\caption{\label{fig1}(a) XRD pattern, (b) SEM image, and (c) elemental mappings of HfMoNbTiZr.}
\end{center}
\end{figure}

The XRD pattern of HfMoNbTiZr can be indexed by the bcc structure with the lattice parameter $a$=3.374(1) \AA, which is determined by the least square method, as shown in Fig.\hspace{1 mm}\ref{fig1}(a).
No extra peaks other than those of the bcc structure are observed.
Figure \ref{fig1}(b) shows a representative SEM image of HfMoNbTiZr, displaying no apparent impurity phase.
The elemental mapping of each element is exhibited in Fig.\hspace{1mm}\ref{fig1}(c), which suggests a homogeneous elemental distribution.
The chemical composition determined by EDX is Hf$_{21.0(5)}$Mo$_{19.8(3)}$Nb$_{20.0(6)}$Ti$_{19.1(5)}$Zr$_{20.1(7)}$, which agrees with the starting composition.

\begin{figure}
\begin{center}
\includegraphics[width=1\linewidth]{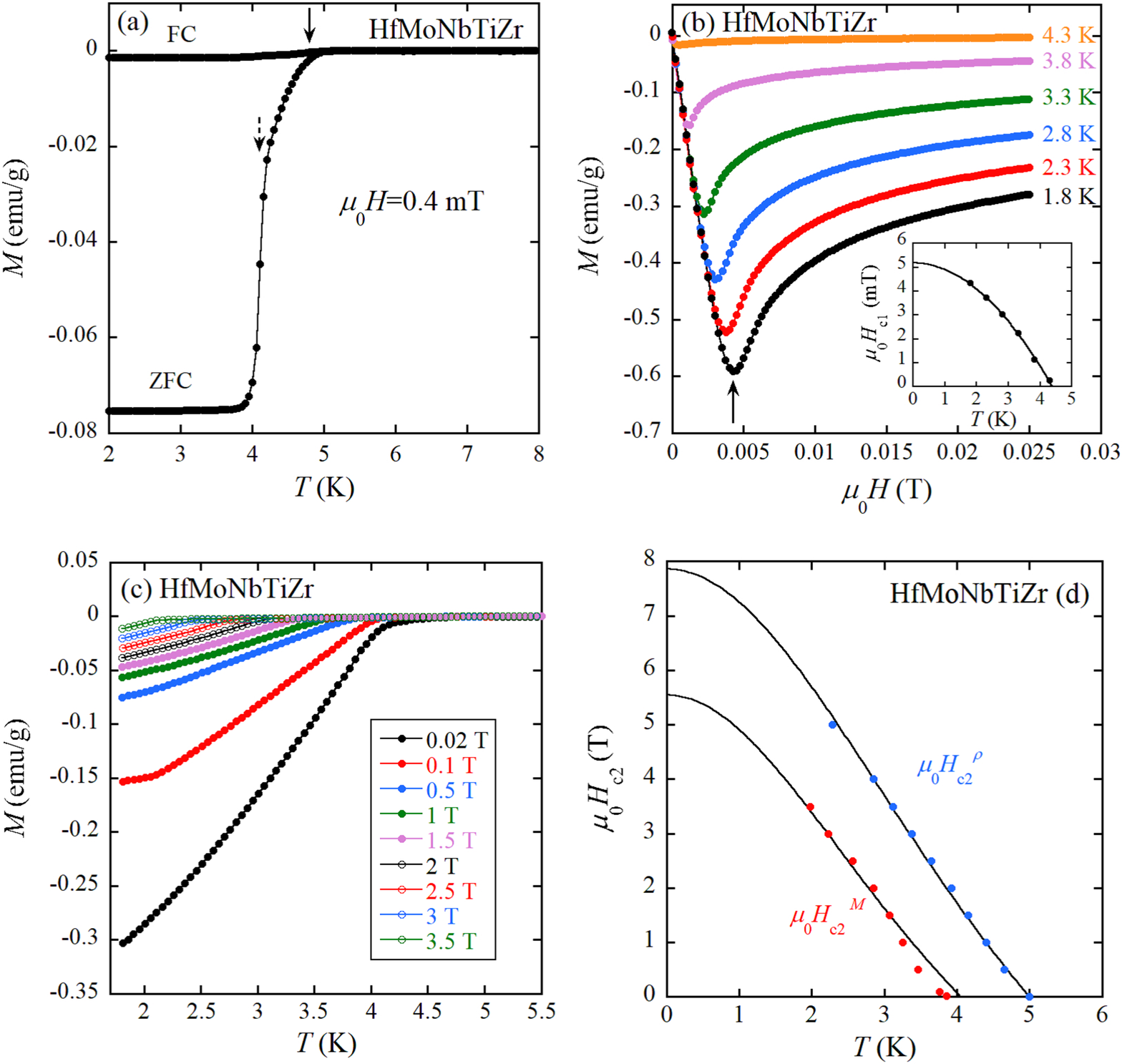}
\caption{\label{fig2}(a) Temperature dependence of $M$ of HfMoNbTiZr. (b) Field-dependent magnetization of HfMoNbTiZr between 1.8 K and 4.3 K. The inset shows the temperature dependence of the lower critical field of HfMoNbTiZr. The solid curve shows the fitting result using Eq. (2). (c) Temperature-dependent ZFC magnetization under the external fields denoted in the figure. (d) Temperature dependences of the upper critical field determined by magnetization ($\mu_{0}H_\mathrm{c2}^{M}$) and electrical resistivity ($\mu_{0}H_\mathrm{c2}^{\rho}$). The solid curves show the fitting results using Eq. (3).}
\end{center}
\end{figure}

\subsection{Superconducting Properties}
$M$($T$) is measured under zero-field cooled (ZFC) and field cooled (FC) conditions, as presented in Fig.\hspace{1mm}\ref{fig2}(a).
The external field $\mu_{0}H$ was 0.4 mT.
The diamagnetic signal is observed in the ZFC data, and the superconducting critical temperature $T_\mathrm{c}$=4.8 K is determined at the point where a difference between the ZFC and FC data appears, as denoted by the solid arrow in Fig.\hspace{1mm}\ref{fig2}(a).
The FC data indicate flux pinning in the sample.
We note that ZFC $M$ shows an inflection at approximately 4.1 K (see the dotted arrow in Fig.\hspace{1mm}\ref{fig2}(a)), which would imply the influence of sample inhomogeneity, as discussed later.
The lower critical field $H_\mathrm{c1}$ was evaluated using the isothermal $M$-$H$ curves collected at different temperatures denoted in Fig.\hspace{1mm}\ref{fig2}(b).
$M$ in the lower field region varies linearly at each temperature and forms a negative peak.
$H_\mathrm{c1}$ is defined at the field showing the negative peak.
The evaluated $H_\mathrm{c1}$ increases with decreasing temperature (see the inset of Fig.\hspace{1mm}\ref{fig2}(b)), which is reproduced well by the Ginzburg-Landau equation as
\begin{equation}
H_\mathrm{c1}(T)=H_\mathrm{c1}(0)\left(1-\left(\frac{T}{T_\mathrm{c}}\right)^{2}\right)
\end{equation}
where $\mu_{0}H_\mathrm{c1}$(0) is determined to be 5.2 mT.

To estimate the upper critical field $H_\mathrm{c2}$, ZFC $M$($T$) was measured under various $\mu_{0}H$ ranging from 0.02 T to 3.5 T (Fig.\hspace{1mm}\ref{fig2}(c)).
The inflection point is defined as $T_\mathrm{c}$, which corresponds to that of the lower $T_\mathrm{c}$ phase.
The temperature showing the inflection is determined by the maximum of the temperature derivative of $M$ and is suppressed as the application of $H$ is increased.
Thus, the obtained $\mu_{0}H_\mathrm{c2}$ vs. $T$ plot is presented in Fig.\hspace{1mm}\ref{fig2}(d) (see $\mu_{0}H_\mathrm{c2}^{M}$).
The temperature dependence can be fitted by the following formula:
\begin{equation}
H_\mathrm{c2}(T)=H_\mathrm{c2}(0)\left(\frac{1-(T/T_\mathrm{c})^{2}}{1+(T/T_\mathrm{c})^{2}}\right).
\end{equation}
The estimated $\mu_{0}H_\mathrm{c2}$(0) is 5.55 T, and the Ginzburg-Landau coherence length $\xi_\mathrm{GL}$(0) is then calculated to be 7.7 nm using the equation:
\begin{equation}
\xi_\mathrm{GL}(0)=\sqrt{\frac{\Phi_{0}}{2\pi\mu_{0}H_\mathrm{c2}(0)}}
\end{equation}
where $\Phi_{0}$ is the magnetic flux quantum (=2.07$\times$10$^{-15}$ Wb).
Furthermore, the magnetic penetration depth $\lambda_\mathrm{GL}$(0) can be extracted from the following relation\cite{He:InorgChem2021}:
\begin{equation}
\mu_{0}H_\mathrm{c1}(0)=\frac{\Phi_{0}}{4\pi\lambda_\mathrm{GL}(0)^{2}}\mathrm{ln}\frac{\lambda_\mathrm{GL}(0)}{\xi_\mathrm{GL}(0)}
\end{equation}
by substituting the $\mu_{0}H_\mathrm{c1}(0)$ and $\xi_\mathrm{GL}$(0) values.
The extracted $\lambda_\mathrm{GL}$(0) is 348 nm.
The Ginzburg-Landau parameter is $\kappa_\mathrm{GL}$=$\lambda_{GL}(0)/\xi_{GL}(0)$=45$\gg$1/$\sqrt{2}$, which indicates that HfMoNbTiZr is a type II superconductor.

\begin{figure}
\begin{center}
\includegraphics[width=0.8\linewidth]{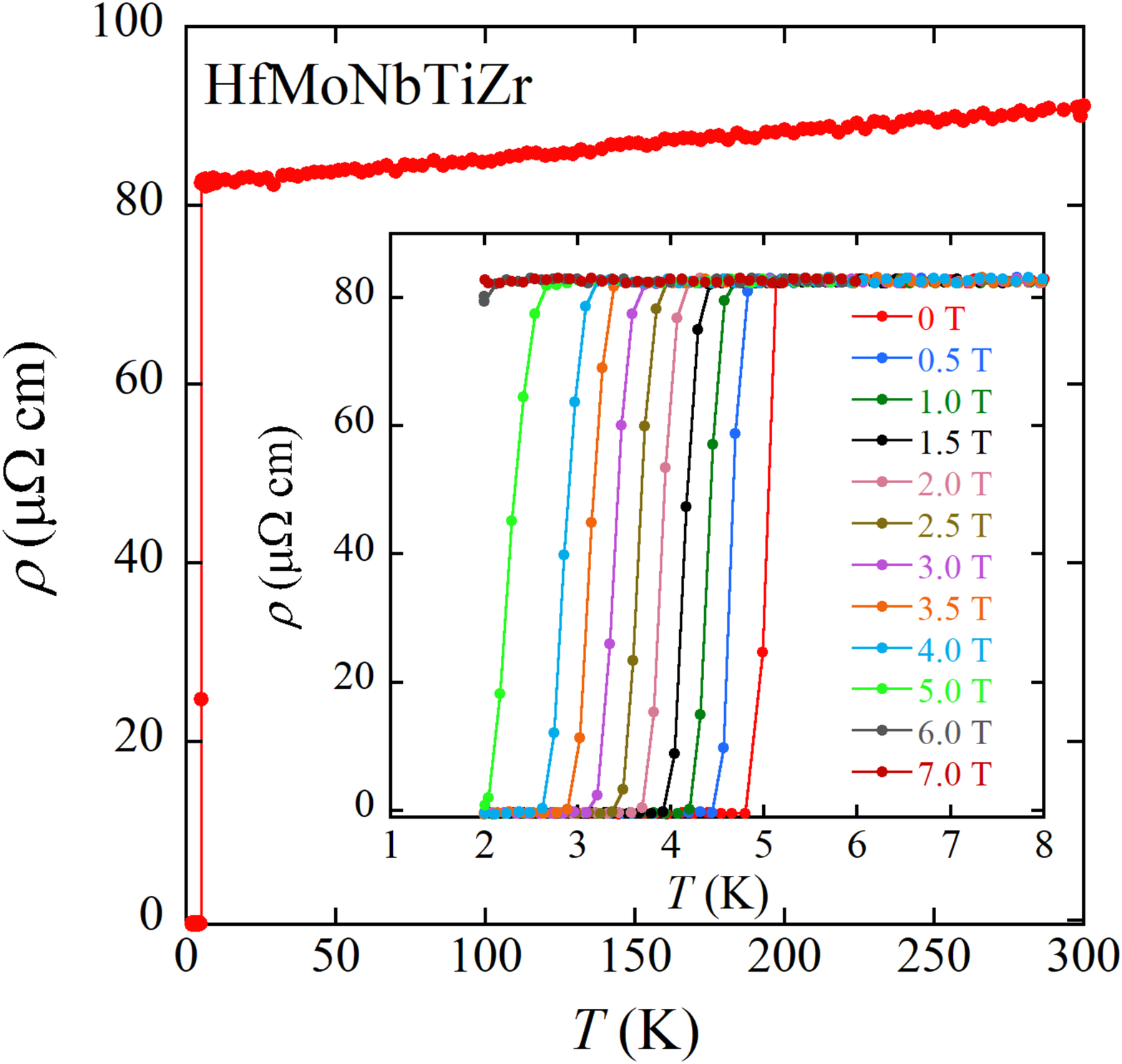}
\caption{\label{fig3}Temperature dependence of electrical resistivity of HfMoNbTiZr at zero field. The inset shows the low-temperature electrical resistivity measured in different external fields (0, 0.5, 1.0, 1.5, 2.0, 2.5, 3.0, 3.5, 4.0, 5.0, 6.0, and 7.0 T).}
\end{center}
\end{figure}

\begin{figure}
\begin{center}
\includegraphics[width=0.8\linewidth]{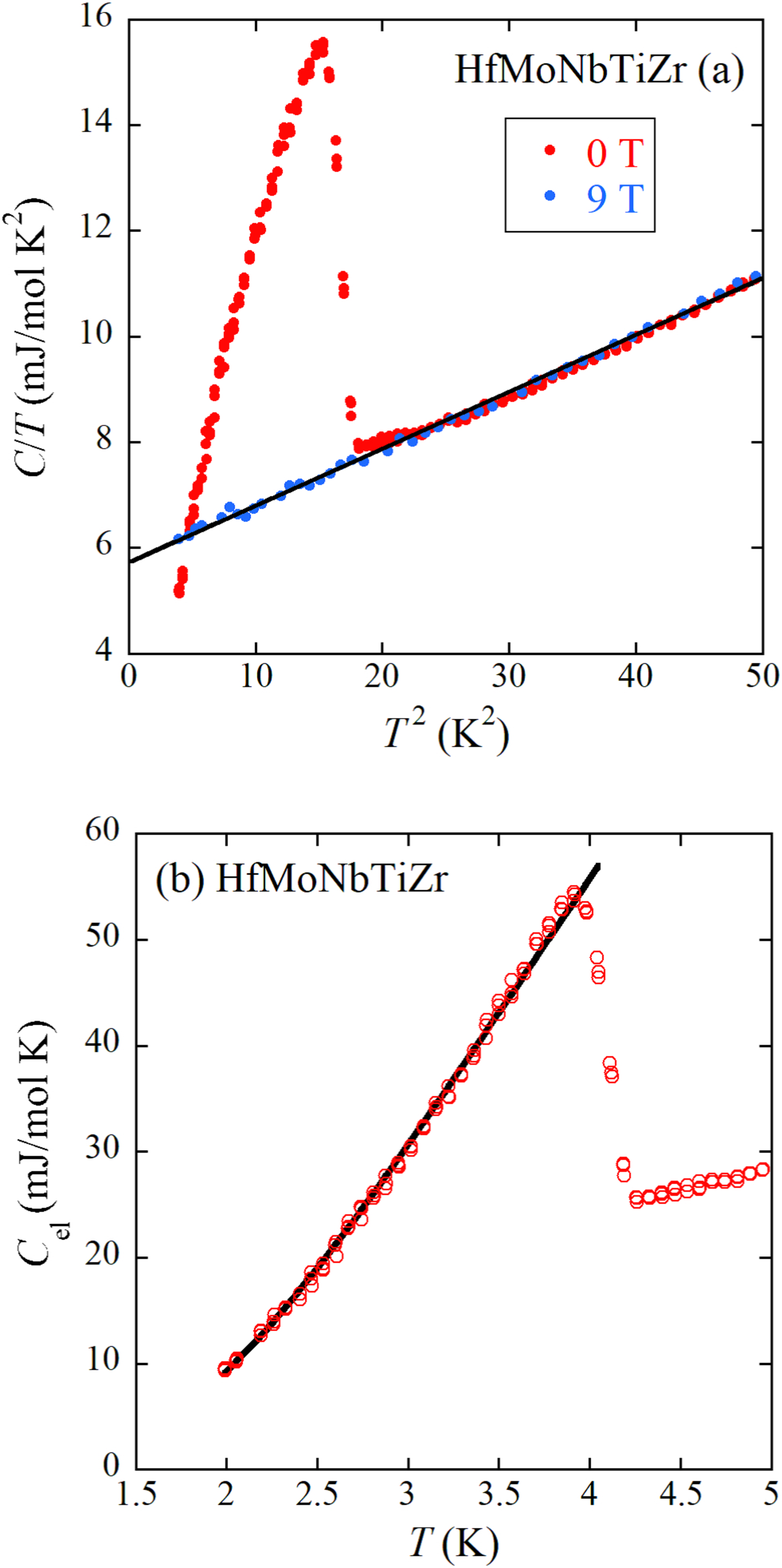}
\caption{\label{fig4}(a) $C/T$ vs. $T^{2}$ plots of HfMoNbTiZr at 0 T and 9 T. The solid line shows the fitted result using $C$($T$)$=\gamma T+\beta T^{3}$ with $\gamma$=5.76 mJ/mol$\cdot$K$^{2}$ and $\beta$=0.1069 mJ/mol$\cdot$K$^{4}$. (b) Low temperature electronic specific heat of HfMoNbTiZr fitted using the BCS model.}
\end{center}
\end{figure}

The normal state $\rho$ exhibits a metallic temperature dependence, which is very weak due to atomic disorders (see Fig.\hspace{1mm}\ref{fig3}).
At the zero field, $\rho$ sharply drops to zero resistivity below approximately 5 K, and $T_\mathrm{c}$ in $\rho$($T$) is defined as the midpoint of the transition ($T_\mathrm{c}$=5 K).
In the specific heat $C$ showing a bulk superconducting transition at 4.1 K (Fig.\hspace{1mm}\ref{fig4}(a)), a very small anomaly appears below 5 K.
Therefore, there may be a parasitic $T_\mathrm{c}$=5 K phase in addition to the main phase with $T_\mathrm{c}$=4.1 K, which is reflected in $M$($T$), showing two anomalies at 4.8 K and 4.1 K.
In HEAs, the high-entropy effect becomes more dominant with increasing temperature, so a perfect solid solution is stable at high temperatures.
However, some HEAs segregate into phases with slightly different chemical compositions after heat treatment\cite{Vrtnik:JALCOM2017,Pacheco:InorgChem2019}.
Thus, while we synthesized the sample through rapid solidification after arc melting and did not carry out annealing, the sample might receive a weak inhomogeneous distribution of chemical compositions during the solidification process.
In such a case, an HEA superconductor can possess a parasitic superconducting phase with slightly different $T_\mathrm{c}$.
We carefully checked the distribution of the chemical composition by EDX, but no noticeable inhomogeneity was detected.
The investigation of sample inhomogeneity will be for performed in future works.
The inset of Fig.\hspace{1mm}\ref{fig3} shows $\rho$($T$) measured at several fields.
$T_\mathrm{c}$ steadily decreases with increasing external fields.
Figure \ref{fig2}(d) also presents the $\mu_{0}H_\mathrm{c2}$ vs. $T$ plot denoted as $\mu_{0}H_\mathrm{c2}^{\rho}$, and the evaluated $\mu_{0}H_\mathrm{c2}$(0) is 7.86 T.
Taking into account the sample inhomogeneity, $H_\mathrm{c2}$(0) determined by $M$ measurements is employed as the intrinsic superconducting parameter.

The results of $C$ measurements are shown in Fig.\hspace{1mm}\ref{fig4}(a), plotting $C/T$ as a function of $T^{2}$.
The sharp discontinuity at $T_\mathrm{c}$=4.1 K is observed, supporting the bulk nature of the superconductive transition.
The transition is depressed under $\mu_{0}H$=9 T, and $C$($T$) can be expressed by
\begin{equation}
C(T)=\gamma T+\beta T^{3}
\end{equation}
where $\gamma$ and $\beta$ are the Sommerfeld coefficient and the lattice contribution, respectively.
These values are determined to be 5.76 mJ/mol$\cdot$K$^{2}$ and 0.1069 mJ/mol$\cdot$K$^{4}$, respectively.
The Debye temperature $\theta_\mathrm{D}$ is derived from the $\beta$ value through the equation:
\begin{equation}
\theta_\mathrm{D}=\left(\frac{12\pi^{4}RN}{5\beta}\right)^{1/3}.
\end{equation}
In this equation, $N$=1 is the number of atoms per formula unit, and the obtained $\theta_\mathrm{D}$ is 263 K.
The electron-phonon coupling $\lambda_\mathrm{e-p}$ can be evaluated using the McMillan formula\cite{Pan:PRB1980}:
\begin{equation}
T_\mathrm{c}=\frac{\theta_\mathrm{D}}{1.45}\mathrm{exp}\left[-\frac{1.04(1+\lambda_\mathrm{e-p})}{\lambda_\mathrm{e-p}-\mu^{*}(1+0.62\lambda_\mathrm{e-p})}\right].
\end{equation}
By employing the Coulomb pseudopotential $\mu^{*}$=0.13, which is widely used for many HEA superconductors, $\theta_\mathrm{D}$=263 K, and $T_\mathrm{c}$=4.1 K, $\lambda_\mathrm{e-p}$ is calculated to be 0.63.

Figure \ref{fig4}(b) displays the temperature dependence of the electronic specific heat $C_\mathrm{el}$($T$), which is obtained by subtracting $\beta T^{3}$ from $C$($T$).
The normalized specific heat jump $\frac{\Delta C_\mathrm{el}}{\gamma T_\mathrm{c}}$ is 1.58, which is close to the BCS value of 1.43 in the weak coupling limit.
In the superconducting state, $C_\mathrm{el}$($T$) can be explained by $a\mathrm{exp}\left(-\frac{\Delta(0)}{k_\mathrm{B}T}\right)$
where $a$ is a proportional constant, $k_\mathrm{B}$ is the Boltzmann constant, and $\Delta(0)$=0.62 meV is the superconducting gap, as shown by the solid curve in Fig.\hspace{1mm}\ref{fig4}(b).
The value of $\frac{2\Delta(0)}{k_\mathrm{B}T_\mathrm{c}}$=3.50 is close to 3.52, which is expected for an s-wave BCS superconductor.

\begin{figure}
\begin{center}
\includegraphics[width=1\linewidth]{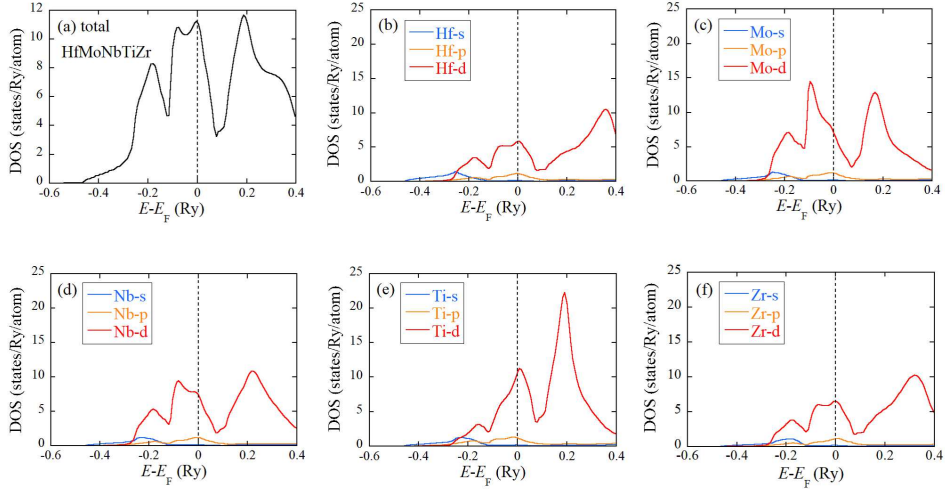}
\caption{\label{fig5}(a) Total electronic density of states for HfMoNbTiZr. (b)-(f) Partial density of states for Hf, Mo, Nb, Ti, and Zr, respectively. The Fermi energy is set to 0 Ry.}
\end{center}
\end{figure}

\begin{figure}
\begin{center}
\includegraphics[width=0.7\linewidth]{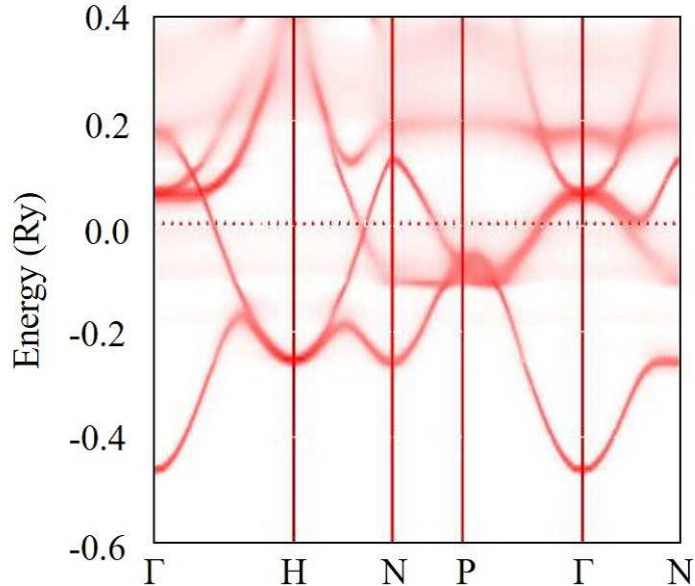}
\caption{\label{fig6}Electronic band structure of HfMoNbTiZr. The Fermi energy is set to 0 Ry.}
\end{center}
\end{figure}

In the electronic structure calculation, the experimental lattice parameter is used.
Figures \ref{fig5}(a) to \hspace{1 mm}\ref{fig5}(f) show the electronic density of states (DOS) of HfMoNbTiZr, indicative of its metallic nature.
The partial DOSs have revealed that d-electron components are dominant at the Fermi level $E_\mathrm{F}$.
In particular, partial DOSs of $d$-electrons in Mo, Nb, and Ti relatively contribute to the total DOS at $E_\mathrm{F}$.
The total DOS at the Fermi level $N(E_\mathrm{F})$ is 11.1 states/Ry/atom.
Using this value and the equation of
\begin{equation}
N(E_\mathrm{F})=\frac{3\gamma_\mathrm{band}}{\pi^{2}k_\mathrm{B}^{2}},
\end{equation}
the Sommerfeld coefficient based on the electronic structure calculation $\gamma_\mathrm{band}$ is 1.92 mJ/mol$\cdot$K$^{2}$, which is smaller than the experimental value (5.76 mJ/mol$\cdot$K$^{2}$).
Considering the mass enhancement due to the electron-phonon coupling, $\gamma$ is described as $\gamma=(1+\lambda_\mathrm{e-p})\gamma_\mathrm{band}$, and $\gamma$ increases to 3.13 mJ/mol$\cdot$K$^{2}$, which is still smaller than the experimental value $\gamma$.
We speculate the origin of further enhancement of $\gamma$ below.
Figure \ref{fig6} shows the electronic band structure of HfMoNbTiZr.
The prominent feature is the broadening of bands caused by atomic disorders.
The band broadening at $E_\mathrm{F}$ reduces $N(E_\mathrm{F})$.
However, as discussed above, some inhomogeneity of the chemical compositions can exist in the actual sample.
$\Delta S_\mathrm{mix}$ of the ideal equiatomic HfMoNbTiZr is the highest in the quinary alloys, which means the highest degree of atomic disorder.
Then, if the composition inhomogeneity partially yields the non-equiatomic Hf-Mo-Nb-Ti-Zr, the lowered degree of atomic disorder suppresses the band broadening at $E_\mathrm{F}$, and $N(E_\mathrm{F})$ increases.
In that case, an enhancement of $\gamma$ is expected.
The other speculation is the mass enhancement due to the electronic correlation.
The Kadowaki-Woods ratio $A/\gamma^{2}$ can evaluate the degree of the many-body effect ($A$: the coefficient of the $T^{2}$ term in $\rho$($T$)).
Many heavy fermion compounds containing Ce or U follow the universal $A/\gamma^{2}$ of 10 $\mu\Omega\cdot$cm$\cdot$mol$^{2}\cdot$K$^{2}$/J$^{2}$.
Recently, it has been reported that HEA superconductors such as Ta$_{1/6}$Nb$_{2/6}$Hf$_{1/6}$Zr$_{1/6}$Ti$_{1/6}$ and Ta$_{34}$Nb$_{33}$Hf$_{8}$Zr$_{14}$Ti$_{11}$ are also located on the line of heavy-fermion compounds, while the $\gamma$ values of HEA superconductors are not so large\cite{Kim:ActaMater2020}.
In our case, the uncertainty of $A$ is large due to the weak temperature dependence of $\rho$ at the normal state.
The evaluation of the $A/\gamma^{2}$ value requires a more precise measurement of $\rho$.

\begin{table}
\caption{\label{tab:table1}%
Superconducting parameters of equiatomic quinary bcc HEA superconductors. Data are from ref.\cite{Vrtnik:JALCOM2017} for HfNbTaTiZr, ref.\cite{Marik:JALCOM2018} for HfNbReTiZr, and ref.\cite{Sarkar:IM2022} for HfNbTaTiV. Each $T_\mathrm{c}$ is determined by specific heat measurement, and accordingly, $\lambda_\mathrm{e-p}$ of HfNbTaTiV is recalculated.}
\begin{tabular}{ccccc}
\hline
Superconducting & HfNbTaTiZr & HfNbReTiZr & HfNbTaTiV & HfMoNbTiZr \\
Parameter & & & & \\
\hline
VEC & 4.4 & 4.8 & 4.6 & 4.6 \\
$T_\mathrm{c}$ (K) & 6.0 & 5.3 & 4.37 & 4.1 \\
$\gamma$ (mJ/mol$\cdot$K$^{2}$) & 7.92 & 5.7 & 8.35 & 5.76 \\
$\theta_\mathrm{D}$ (K) & 212 & 255 & 225 & 263 \\
$\lambda_\mathrm{e-p}$ & 0.77 & 0.69 & 0.67 & 0.63 \\
$\mu_{0}H_\mathrm{c1}$ (mT) & - & 33 & 19.5 & 5.2 \\
$\mu_{0}H_\mathrm{c2}$ (T) & 10.45 & 8.88 & 6.63 & 5.55 \\
$\xi_\mathrm{GL}$ (nm) & 5.6 & 6.1 & 7.1 & 7.7 \\
\hline
\end{tabular}
\end{table}

\begin{figure}
\begin{center}
\includegraphics[width=1\linewidth]{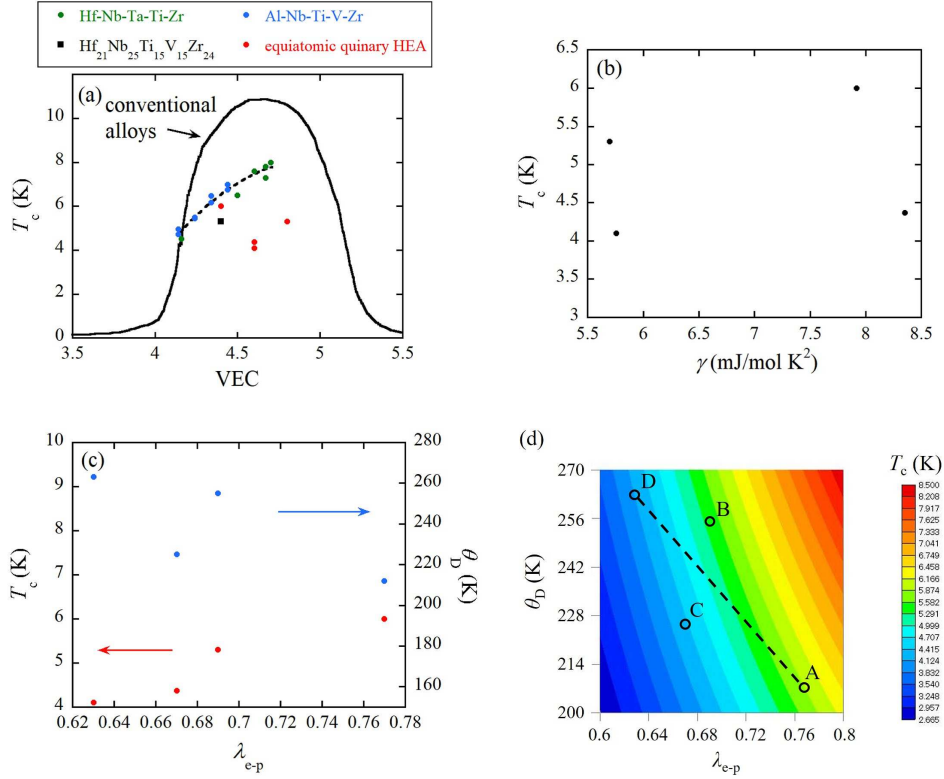}
\caption{\label{fig7}(a) VEC dependence of $T_\mathrm{c}$ of typical non-equiatomic and equiatomic quinary bcc HEA superconductors. The solid line represents the Matthias rule of conventional binary or ternary transition metal alloys. The dotted line is a guide. Relationship between (b) $T_\mathrm{c}$ and $\gamma$ and (c) $T_\mathrm{c}$ (or $\theta_\mathrm{D}$) and $\lambda_\mathrm{e-p}$ for equiatomic quinary bcc HEA superconductors. (d) $T_\mathrm{c}$ contour map as a function of $\theta_\mathrm{D}$ and $\lambda_\mathrm{e-p}$ (A: HfNbTaTiZr, B: HfNbReTiZr, C: HfNbTaTiV, and D: HfMoNbTiZr). The dotted line is a guide for the eyes.}
\end{center}
\end{figure}

Here, we discuss the comparison of superconducting properties among equiatomic quinary bcc HEA superconductors.
The major superconducting parameters are summarized in Table \ref{tab:table1}.
Figure \ref{fig7}(a) presents the VEC dependence of $T_\mathrm{c}$ of equiatomic quinary HEA superconductors and several typical non-equiatomic superconductors\cite{Ishizu:RINP,Harayama:JSNM2021,Rohr:PNAS2016}.
As shown by the dotted line, the VEC dependence of $T_\mathrm{c}$ for many non-equiatomic quinary HEA superconductors displays a trend similar to the Matthias rule of conventional binary or ternary transition metal alloys (see the solid curve), reflecting the vital role of $N(E_\mathrm{F})$ for the determination of $T_\mathrm{c}$, considering that VEC is related to the information of $N(E_\mathrm{F})$.
However, the trend of equiatomic quinary bcc HEA superconductors seems to contradict the Matthias rule; $T_\mathrm{c}$ decreases even with increasing VEC above 4.4.
The reason is that factors other than $N(E_\mathrm{F})$ dominate in determining $T_\mathrm{c}$.
This is also supported by the $T_\mathrm{c}$ vs. $\gamma$ plot showing no clear systematic behavior in Fig.\hspace{1mm}\ref{fig7}(b).
We have investigated the effect of electron-phonon coupling on $T_\mathrm{c}$'s of equiatomic quinary bcc HEA superconductors, as shown in Fig.\hspace{1mm}\ref{fig7}(c), clearly displaying the growth of $T_\mathrm{c}$ with increasing $\lambda_\mathrm{e-p}$.
Moreover, $\lambda_\mathrm{e-p}$ seems to be negatively correlated with $\theta_\mathrm{D}$, which is contrary to our expectation; a compound with a higher $\theta_\mathrm{D}$ tends to possess an enhanced $\lambda_\mathrm{e-p}$, and a higher $T_\mathrm{c}$ is achieved.
It should be noted that each $\lambda_\mathrm{e-p}$ is calculated using the same $\mu^{*}$ value of 0.13. The phenomenological parameter $\mu^{*}$ varies from alloy to alloy, and the relation between $T_\mathrm{c}$ and $\lambda_\mathrm{e-p}$ might be broken. However, considering that $T_\mathrm{c}$ mainly depends on $N(E_\mathrm{F})$ and the electron-phonon interaction and that $N(E_\mathrm{F})$ is not the decisive factor for $T_\mathrm{c}$ in the present case, the electron-phonon interaction plays a vital role. Therefore, the electron-phonon interaction would be weakened as $T_\mathrm{c}$ is lowered, and the relation between $T_\mathrm{c}$ and $\lambda_\mathrm{e-p}$ presented in Fig.\hspace{1mm}\ref{fig7}(c) cannot be far from the true value in the four equiatomic HEAs.

The mechanism behind the relation between $\theta_\mathrm{D}$ and $T_\mathrm{c}$ in equiatomic quinary bcc HEA superconductors can be explained as follows.
As shown in Fig.\hspace{1mm}\ref{fig6}, the electronic band structure shows band broadening due to atomic disorders.
The phonon band structure is also expected to show phonon broadening, which would be more significant with increasing phonon frequency.
We note that such phonon broadening in HEAs has been theoretically investigated\cite{Kormann:npjCM2017,Ikeda:MC2019}.
Therefore, a bcc HEA with a higher $\theta_\mathrm{D}$ would possess a broader phonon band, indicating a shorter phonon lifetime, considering the uncertainty principle $\Delta E\Delta t\geq\hbar/2$, where $\Delta E$ is the energy uncertainty associated with the band broadening, $\Delta t$ is the lifetime, and $\hbar$ is the reduced Planck's constant.
The shorter lifetime caused by a higher $\theta_\mathrm{D}$ leads to weakened electron-phonon coupling.
To check the effect of $\theta_\mathrm{D}$ and $\lambda_\mathrm{e-p}$ on $T_\mathrm{c}$ in Eq. (8), we calculated $T_\mathrm{c}$ as a function of $\theta_\mathrm{D}$ and $\lambda_\mathrm{e-p}$ and made the $T_\mathrm{c}$ contour map in Fig.\hspace{1mm}\ref{fig7}(d).
The four equiatomic HEAs are plotted (A: HfNbTaTiZr, B: HfNbReTiZr, C: HfNbTaTiV, and D: HfMoNbTiZr).
Going from A to D along the dotted guideline with increasing $\theta_\mathrm{D}$ and decreasing $\lambda_\mathrm{e-p}$, $T_\mathrm{c}$ is systematically lowered, which supports the reduction of $T_\mathrm{c}$ by increasing $\theta_\mathrm{D}$.
In Eq. (8), $T_\mathrm{c}$ increases as $\theta_\mathrm{D}$ is increased under no influence of the exponential term.
The exponential term reflects the strength of electron-phonon coupling and is reduced with decreasing $\lambda_\mathrm{e-p}$.
Therefore, if the reduction of $T_\mathrm{c}$ by weakened $\lambda_\mathrm{e-p}$ dominates over the enhancement of $T_\mathrm{c}$ by increasing $\theta_\mathrm{D}$, the resultant $T_\mathrm{c}$ is lowered.
Here, we comment that the observation of the decrease in $T_\mathrm{c}$ with increasing $\theta_\mathrm{D}$ is now restricted in the four equiatomic HEAs, and further studies are required to conclude whether the relationship can be generalized in all equiatomic HEA superconductors.

\begin{table}
\caption{\label{tab:table2}%
Vickers microhardness, VEC, and $T_\mathrm{c}$ of bcc HEA superconductors.}
\begin{tabular}{cccc}
\hline
Alloy & Vickers microhardness & VEC & $T_\mathrm{c}$ \\
      & (HV) & & (K)  \\
\hline
HfMoNbTiZr & 398(5) & 4.6 & 4.1 \\
Hf$_{21}$Nb$_{25}$Ti$_{15}$V$_{15}$Zr$_{24}$ & 389(5) & 4.4 & 5.3 \\
(TaNb)$_{0.7}$(ZrHfTi)$_{0.3}$ & 336(5) & 4.7 & 8.0 \\
Al$_{5}$Nb$_{24}$Ti$_{40}$V$_{5}$Zr$_{26}$ & 352(10) & 4.24 & 5.5 \\
Al$_{5}$Nb$_{14}$Ti$_{35}$V$_{5}$Zr$_{41}$ & 282(5) & 4.14 & 4.73 \\
Al$_{5}$Nb$_{24}$Ti$_{35}$V$_{5}$Zr$_{31}$ & 325(2) & 4.24 & 5.45 \\
Al$_{5}$Nb$_{34}$Ti$_{35}$V$_{5}$Zr$_{21}$ & 320(4) & 4.34 & 6.47 \\
Al$_{5}$Nb$_{44}$Ti$_{35}$V$_{5}$Zr$_{11}$ & 295(10) & 4.44 & 6.75 \\
Al$_{5}$Nb$_{14}$Ti$_{45}$V$_{5}$Zr$_{31}$ & 262(9) & 4.14 & 4.96 \\
Al$_{5}$Nb$_{34}$Ti$_{25}$V$_{5}$Zr$_{31}$ & 323(6) & 4.34 & 6.18 \\
Al$_{5}$Nb$_{44}$Ti$_{15}$V$_{5}$Zr$_{31}$ & 350(5) & 4.44 & 6.98 \\
\hline
\end{tabular}
\end{table}

\begin{figure}
\begin{center}
\includegraphics[width=0.8\linewidth]{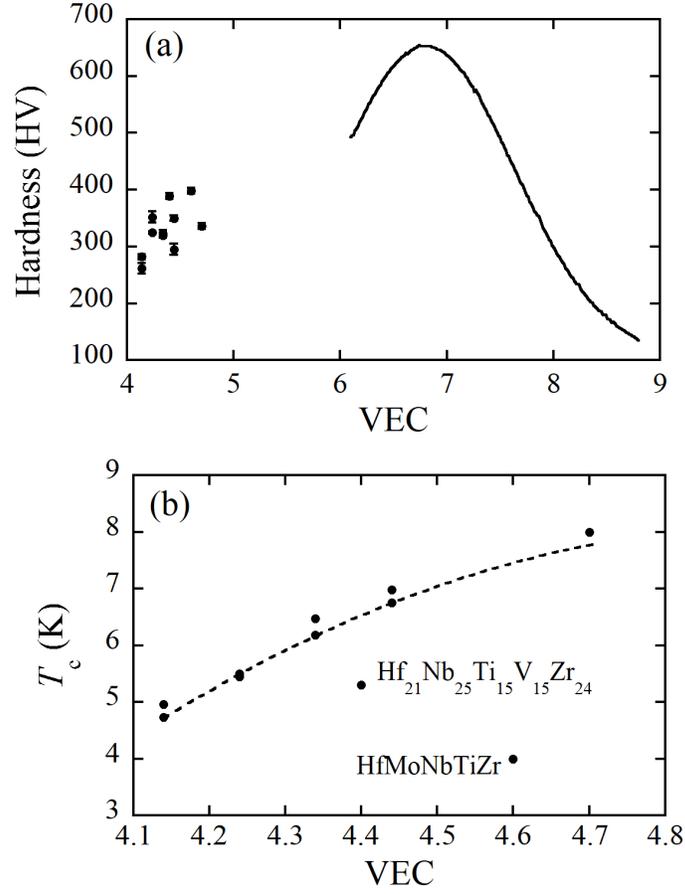}
\caption{\label{fig8}(a) VEC dependence of Vickers microhardness of bcc HEA superconductors investigated. The solid curve represents the guideline obtained for non-superconducting HEAs with bcc structures (VEC: 6.0$\sim$7.55) and fcc structures (VEC: 7.8$\sim$8.8) taken from ref.\cite{Tian:IM2015}. (b) VEC dependence of $T_\mathrm{c}$ of bcc HEA superconductors used for Vickers microhardness examinations. The dotted line is a guide.}
\end{center}
\end{figure}

\subsection{Hardness}
Table II shows the results of Vickers microhardness measurements for bcc HEA superconductors reported by our team\cite{Ishizu:RINP,Harayama:JSNM2021}.
We carried out measurements for specimens with various VEC values because it is proposed that the Vickers microhardness depends on the VEC.
The VEC dependence of hardness may reflect the fact that hard elements are prone to locate on VEC: 5$\sim$7.
Figure \ref{fig8}(a) exhibits the VEC dependence of hardness, with a guideline obtained for non-superconducting HEAs with the bcc structures (VEC: 6.0$\sim$7.55) and fcc structures (VEC: 7.8$\sim$8.8) described in ref.\cite{Tian:IM2015}.
The hardness of bcc HEA superconductors tends to increase with increasing VEC, and the plot might be connected to the guideline, indicating the universal relation between the VEC and the hardness.
In addition, a deep learning study of the hardness of refractory HEAs with the VEC ranging from 4 to 6 also indicates that the hardness increases as the VEC is increased\cite{Bhandari:Crystals2021}.

In Figure \ref{fig8}(b), the dotted line represents the VEC dependence of $T_\mathrm{c}$ for typical quinary bcc HEA superconductors, which are mostly non-equiatomic.
The data points of HfMoNbTiZr and Hf$_{21}$Nb$_{25}$Ti$_{15}$V$_{15}$Zr$_{24}$ deviate from the dotted line, and the hardness of these alloys are relatively higher.
This might suggest that the high hardness in bcc HEA superconductors reduces $T_\mathrm{c}$.
In equiatomic quinary bcc HEA superconductors other than HfMoNbTiZr, the hardness of HfNbTaTiZr with no deformation is reported to be 295 HV\cite{Lin:JALCOM2015}, which is lower than 398 HV of HfMoNbTiZr, which also supports the decrease in $T_\mathrm{c}$ with increasing hardness.
Considering that $\theta_\mathrm{D}$ generally reflects the hardness of a material, the mechanism behind the relation between $\theta_\mathrm{D}$ and $T_\mathrm{c}$ also works in the hardness dependence of $T_\mathrm{c}$.
According to Fig.\hspace{1mm}\ref{fig8}(a), the hardness of the bcc HEA increases with the VEC.
Therefore, it may be difficult to achieve a high $T_\mathrm{c}$ bcc HEA superconductor simultaneously showing high hardness.

The hardness of HEA superconductors is reported in (MoReRu)$_{(1-2x)/3}$(PdPt)$_{x}$C$_{y}$ with hcp and fcc structures\cite{Zhu:JALCOM:2022}.
While the C-free sample with 0.042$\leq x \leq$0.167 forms the hcp structure, C-doping induces a structural transformation into the fcc structure.
In this system, the covalent bonding between the metal and carbon atoms leads to a Vickers microhardness exceeding 1000 HV.
The VEC of (MoReRu)$_{(1-2x)/3}$(PdPt)$_{x}$C$_{y}$ ranges from 6.25 to 8.0, and the Vickers microhardness continuously increases with decreasing VEC.
This behavior qualitatively agrees with the result in Fig.\hspace{1mm}\ref{fig8}(a).
$T_\mathrm{c}$ also systematically depends on the VEC in each crystal structure; $T_\mathrm{c}$ increases with decreasing VEC.
Therefore, in (MoReRu)$_{(1-2x)/3}$(PdPt)$_{x}$C$_{y}$, the enhancement of Vickers microhardness seems to increase $T_\mathrm{c}$, which is contrasted with the results of bcc HEA superconductors.

\section{Summary}
We have found that bcc HfMoNbTiZr is the type-II BCS HEA superconductor with $T_\mathrm{c}$=4.1 K.
The lower and upper critical fields are 5.2 mT and 5.82 T, respectively.
The $\gamma$ and $\theta_\mathrm{D}$ derived from the specific heat are 5.76 mJ/mol$\cdot$K$^{2}$ and 263 K, respectively.
$\lambda_\mathrm{e-p}$ was calculated to be 0.63.
The electronic structure calculation revealed that partial DOSs of $d$-electrons in Mo, Nb, and Ti are relatively dominant at $E_\mathrm{F}$.
Furthermore, the calculation suggests band broadening due to atomic disorder.
The superconducting properties are compared among equiatomic quinary bcc HEA superconductors.
$T_\mathrm{c}$ decreases with decreasing $\lambda_\mathrm{e-p}$, which is negatively correlated with $\theta_{D}$.
The mechanism behind the relation between $\lambda_\mathrm{e-p}$ and $\theta_{D}$ would be explained by a shorter phonon lifetime at a higher $\theta_{D}$, which is based on phonon broadening due to atomic disorders and the uncertainty principle.
The hardness of refractory bcc HEA superconductors tends to increase with increasing VEC.
$T_\mathrm{c}$ is exceptionally low in the bcc HEA superconductor with relatively high hardness, which would be understood by considering that the hardness generally increases with increasing $\theta_\mathrm{D}$, and $\theta_\mathrm{D}$ is negatively correlated with $T_\mathrm{c}$ in the high-entropy state.

\section*{CRediT authorship contribution statement}
Jiro Kitagawa: Supervision, Formal analysis, Writing - original draft, Writing - reviewing \& editing. Kazuhisa Hoshi: Investigation. Yuta Kawasaki: Investigation. Rikuo Koga: Investigation. Yoshikazu Mizuguchi: Investigation, Formal analysis, Writing - reviewing \& editing. Terukazu Nishizaki: Investigation, Formal analysis, Writing - reviewing \& editing.

\section*{Acknowledgments}
J.K. is grateful for the support provided by the Comprehensive Research Organization of Fukuoka Institute of Technology. Y.M. acknowledges the support from a Grant-in-Aid for Scientific Research (KAKENHI) (Grant No. 21H00151). T.N. acknowledges the support from a Grant-in-Aid for Scientific Research (KAKENHI) (Grant No. 20K03867) and the Advanced Instruments Center of Kyushu Sangyo University.

\section*{Declaration of Competing Interest}
The authors have no conflicts to disclose.


\begin{thebibliography}{99}
\bibitem{Wang:JMCA2021}
X. Wang, W. Guo, Y. Fu, J. Mater. Chem. A 9 (2021) 663.

\bibitem{Sathiyamoorthi:PMS2022}
P. Sathiyamoorthi, H. S. Kim, Prog. Mater. Sci. 123 (2022) 100709.

\bibitem{Chaudhary:MT2021}
V. Chaudhary, R. Chaudhary, R. Banerjee, R. V. Ramanujan, Mater. Today 49 (2021) 231.

\bibitem{Musico:APLMater2020}
B. L. Music\'{o}, D. Gilbert, T. Z. Ward, K. Page, E. George, J. Yan, D. Mandrus, V. Keppens, APL Mater. 8 (2020) 040912.

\bibitem{Ying:JACS}
T. Ying, T. Yu, Y.-S. Shiah, C. Li, J. Li, Y. Qi, H. Hosono, J. Am. Chem. Soc. 143 (2021) 7042.

\bibitem{Iwan:AIPAdv2021}
S. Iwan, K. C. Burrage, B. C. Storr, S. A. Catledge, Y. K. Vohra, R. Hrubiak, N. Velisavljevic, AIP Adv. 11 (2021) 035107.

\bibitem{Wang:AAC2022}
Y. Wang, Adv. Appl. Ceram. 121 (2022) 57.

\bibitem{Lu:ASS2021}
X. Lu, C. Zhang, C. Wang, X. Cao, R. Ma, X. Sui, J. Hao, W. Liu, Appl. Surf. Sci. 557 (2021) 149813.

\bibitem{Li:PMS2021}
W. Li, D. Xie, D. Li, Y. Zhang, Y. Gao, P. K. Liaw, Prog. Mater. Sci.118 (2021) 100777.

\bibitem{Marques:EES2021}
F. Marques, M. Balcerzak, F. Winkelmann, G. Zepon, M. Felderhoff, Energy Environ. Sci. 14 (2021) 5191.

\bibitem{Pickering:Entropy2021}
E. J. Pickering, A. W. Carruthers, P. J. Barron, S. C. Middleburgh, D. E. J. Armstrong, A. S. Gandy, Entropy, 23 (2021) 98.

\bibitem{Castro:Metals2021}
D. Castro, P. Jaeger, A. C. Baptista, J. P. Oliveira, Metals 11 (2021) 648.

\bibitem{Wu:JACS2020}
D. Wu, K Kusada, T. Yamamoto, T. Toriyama, S. Matsumura, S. Kawaguchi, Y. Kubota, H. Kitagawa, J. Am. Chem. Soc. 142 (2020) 13833.

\bibitem{Jiang:Science2021}
B. Jiang, Y. Yu, J. Cui, X. Liu, L. Xie, J. Liao, Q. Zhang, Y. Huang, S. Ning, B. Jia, B. Zhu, S. Bai, L. Chen, S. J. Pennycook, J. He, Science 371 (2021) 830.

\bibitem{Marik:ScrMater2020}
S. Marik, D. Singh, B. Gonano, F. Veillon, D. Pelloquin, Y. Br\'{e}ard, Scr. Mater. 186 (2020) 366.

\bibitem{Yang:NM2022}
B. Yang, Y. Zhang, H. Pan, W. Si, Q. Zhang, Z. Shen, Y. Yu, S. Lan, F. Meng, Y. Liu, H. Huang, J. He, L. Gu, S. Zhang, L. -Q. Chen, J. Zhu, C. -W. Nan, Y. -H. Lin, Nat. Mater. (2022) https://doi.org/10.1038/s41563-022-01274-6.

\bibitem{Bhardwaj:TI2021}
V. Bhardwaj, Q. Zhou, F. Zhang, W. Han, Y. Du, K. Hua, H. Wang, Tribol. Int. 160 (2021) 107031.

\bibitem{Rao:AFM2021}
Z. Rao, B. Dutta, F. K\"{o}rmann, W. Lu, X. Zhou, C. Liu, A. Kwiatkowski da Silva, U. Wiedwald, M. Spasova, M. Farle, D. Ponge, B. Gault, J. Neugebauer, D. Raabe, Z. Li, Adv. Funct. Mater. 31 (2021) 2007668.

\bibitem{Zherebtsov:Int}
S. Zherebtsov, N. Yurchenko, E. Panina, M. Tikhonovsky, N. Stepanov, Intermetallics 116 (2020) 106652.

\bibitem{Kitagawa:APLMater2022}
J. Kitagawa, M. Fukuda, S. Fukuda, K. Fujiki, Y. Nakamura, T. Nishizaki, APL Mater. 10 (2022) 071101.

\bibitem{Kozelj:PRL2014}
P. Ko\v{z}elj, S. Vrtnik, A. Jelen, S. Jazbec, Z. Jagli\v{c}i\'{c}, S. Maiti, M. Feuerbacher, W. Steurer, J. Dolin\v{s}ek, Phys. Rev. Lett. 113 (2014) 107001.

\bibitem{Sun:PRM2019}
L. Sun, R.J. Cava, Phys. Rev. Mater. 3 (2019) 090301.

\bibitem{Kitagawa:Metals2020}
J. Kitagawa, S. Hamamoto, N. Ishizu, Metals 10 (2020) 1078.

\bibitem{Rohr:PRM2018}
F. O. von Rohr, R. J. Cava, Phys. Rev. Mater. 2 (2018) 034801.

\bibitem{Marik:JALCOM2018}
S. Marik, M. Varghese, K. P. Sajilesh, D. Singh, R. P. Singh, J. Alloys Compd. 769 (2018) 1059.

\bibitem{Ishizu:RINP}
N. Ishizu, J. Kitagawa, Res. Phys. 13 (2019) 102275.

\bibitem{Harayama:JSNM2021}
Y. Harayama, J. Kitagawa, J. Supercond. Nov. Magn. 34 (2021) 2787.

\bibitem{Sarkar:IM2022}
N. K. Sarkar, C. L. Prajapat, P. S. Ghosh, N. Garg, P. D. Babu, S. Wajhal, P. S. R. Krishna, M. R. Gonal, R. Tewari, P. K. Mishra, Intermetallics 144 (2022) 107503.

\bibitem{Motla:PRB2022}
K. Motla, P. K. Meena, Arushi, D. Singh, P. K. Biswas, A. D. Hillier, R. P. Singh, Phys. Rev. B 105 (2022) 144501.

\bibitem{Lee:PhysicaC2019}
Y. -S. Lee, R. J. Cava, Physica C 566 (2019) 1353520.

\bibitem{Marik:PRM2019}
S. Marik, K. Motla, M. Varghese, K. P. Sajilesh, D. Singh, Y. Breard, P. Boullay, R. P. Singh, Phys. Rev. Mater. 3 (2019) 060602(R).

\bibitem{Zhu:JALCOM:2022}
Q. Zhu, G. Xiao, Y. Cui, W. Yang, S. Song, G. -H. Cao, Z. Ren, J. Alloys Compd. 909 (2022) 164700.

\bibitem{Stolze:ChemMater2018}
K. Stolze, J. Tao, F. O. von Rohr, T. Kong, R. J. Cava, Chem. Mater. 30 (2018) 906.

\bibitem{Wu:SCM2020}
J. Wu, B. Liu, Y. Cui, Q. Zhu, G. Xiao, H. Wang, S. Wu, G. Cao, Z. Ren, Sci. China Mater. 63 (2020) 823.

\bibitem{Yamashita:JALCOM2021}
A. Yamashita, T. D. Matsuda, Y. Mizuguchi, J. Alloys Compd. 868 (2021) 159233.

\bibitem{Mizuguchi:JPSJ2019}
Y. Mizuguchi, J. Phys. Soc. Jpn. 88 (2019) 124708.

\bibitem{Yamashita:DalTran2020}
A. Yamashita, R. Jha, Y. Goto, T. D. Matsuda, Y. Aokia, Y. Mizuguchi, Dalton Trans. 49 (2020) 9118.

\bibitem{Stolze:JMCC2018}
K. Stolze, F. A. Cevallos, T. Kong, R. J. Cava, J. Mater. Chem. C 6 (2018) 10441.

\bibitem{Liu:ACS2020}
B. Liu, J. Wu, Y. Cui, Q. Zhu, G. Xiao, H. Wang, S. Wu, G. Cao, Z. Ren, ACS Appl. Electron. Mater. 2 (2020) 1130.

\bibitem{Kasen:SST2021}
M. R. Kasem, A. Yamashita, T. Hatano, K. Sakurai, N. Oono-Hori, Y. Goto, O. Miura, Y. Mizuguchi, Supercond. Sci. Technol. 34 (2021) 125001.

\bibitem{Sogabe:SSC2019}
R. Sogabe, Y. Goto, T. Abe, C. Moriyoshi, Y. Kuroiwa, A. Miura, K. Tadanaga, Y. Mizuguchi, Solid State Commun. 295 (2019) 43.

\bibitem{Shukunami:PhysicaC2020}
Y. Shukunami, A. Yamashita, Y. Goto, Y. Mizuguchi, Physica C 572 (2020) 1353623.

\bibitem{Guo:PNAC2017}
J. Guo, H. Wang, F. von Rohr, Z. Wang, S. Cai, Y. Zhou, K. Yang, A. Li, S. Jiang, Q. Wu, R. J. Cava, L. Sun, Proc. Natl. Acad. Sci. 114 (2017) 13144.

\bibitem{Liu:JALCOM2021}
B. Liu, J. Wu, Y. Cui, Q. Zhu, G. Xiao, S. Wu, G.-h. Cao, Z. Ren, J. Alloys Compd. 869 (2021) 159293.

\bibitem{Zhang:PRR2020}
X. Zhang, N. Winter, C. Witteveen, T. Moehl, Y. Xiao, F. Krogh, A. Schilling, F. O. von Rohr, Phys. Rev. Res. 2 (2020) 013375.

\bibitem{Shu:APL2022}
R. Shu, X. Zhang, S. Rao, A. Febvrier, P. Eklund, Appl. Phys. Lett. 120 (2022) 151901.

\bibitem{Gao:APL2022}
L. Gao, T. Ying, Y. Zhao, W. Cao, C. Li, L. Xiong, Q. Wang, C. Pei, J.-Y. Ge, H. Hosono, Y. Qi, Appl. Phys. Lett. 120 (2022) 092602.

\bibitem{Kim:ActMater2022}
J. H. Kim, R. Hidayati, S. -G. Jung, Y. A. Salawu, H. -J. Kim, J. H. Yun, J. -S. Rhyee, Acta Mater. 232 (2022) 117971.

\bibitem{Jung:NC2022}
S. -G. Jung, Y. Han, J. H. Kim, R. Hidayati, J. -S. Rhyee, J. M. Lee, W. N. Kang, W. S. Choi, H. -R. Jeon, J. Suk, T. Park, Nat. Commun. 13 (2022) 3373.

\bibitem{Rohr:PNAS2016}
F. von Rohr, M. J. Winiarski, J. Tao, T. Klimczuk, R. J. Cava, Proc. Natl. Acad. Sci. 113 (2016) E7144.

\bibitem{Vrtnik:JALCOM2017}
S. Vrtnik, P. Ko\v{z}elj, A. Meden, S. Maiti, W. Steurer, M. Feuerbacher, J. Dolin\v{s}ek, J. Alloys Compd. 695 (2017) 3530.

\bibitem{Guo:MD2015}
N. N. Guo, L. Wang, L. S. Luo, X. Z. Li, Y. Q. Su, J. J. Guo, H. Z. Fu, Mater. Des. 81 (2015) 87.

\bibitem{Tseng:Entropy2019}
K. -K. Tseng, C. -C. Juan, S. Tso, H. -C. Chen, C. -W. Tsai, J. -W. Yeh, Entropy 21 (2019) 15.

\bibitem{Ge:ICF2019}
Y. Ge, S. Ma, K. Bao, Q. Tao, X. Zhao, X. Feng, L. Li, B. Liu, P. Zhu, T. Cui, Inorg. Chem. Front. 6 (2019) 1282.

\bibitem{Cui:JALCOM2021}
Y. Cui, J. Wu, B. Liu, Q. Zhu, G. Xiao, S. Wu, G. Cao, and Z. Ren, J. Alloys Compd. 856 (2021) 157314.

\bibitem{Akai:JPSJ1982}
H. Akai, J. Phys. Soc. Jpn. 51 (1982) 468.

\bibitem{Fukushima:PRM2022}
T. Fukushima, H. Akai, T. Chikyow, H. Kino, Phys. Rev. Mater. 6 (2022) 023802.

\bibitem{He:InorgChem2021}
Z. He, R. Huang, K. Zhou, Y. Liu, S. Guo, Y. Song, Z. Guo, S. Hu, L. He, Q. Huang, L. Li, J. Zhang, S. Wang, J. Guo, X. Xing, J. Chen, Inorg. Chem. 60 (2021) 6157.

\bibitem{Pacheco:InorgChem2019}
V. Pacheco, G. Lindwall, D. Karlsson, J. Cedervall, S. Fritze, G. Ek, P. Berastegui, M. Sahlberg, U. Jansson, Inorg. Chem. 58 (2019) 811.

\bibitem{Pan:PRB1980}
P. H. Pan, D. K. Finnemore, A. J. Bevolo, H. R. Shanks, B. J. Beaudry, F. A. Schmidt, G. C. Danielson, Phys. Rev. B 21 (1980) 2809.

\bibitem{Kim:ActaMater2020}
G. Kim, M. -H. Lee, J. H. Yun, P. Rawat, S. -G. Jung, W. Choi, T. -S. You, S. J. Kim, J. -S. Rhyee,  Acta Mater. 186 (2020) 250.

\bibitem{Kormann:npjCM2017}
F. K\"{o}rmann, Y. Ikeda, B. Grabowski, M. H. F. Sluiter, Npj Comput. Mater. 3 (2017) 36.

\bibitem{Ikeda:MC2019}
Y. Ikeda, B. Grabowski, F. K\"{o}rmann, Mater. Charact. 147 (2019) 464.

\bibitem{Tian:IM2015}
F. Tian, L. K. Varga, N. Chen, J. Shen, L. Vitos, Intermetallics 58 (2015) 1.

\bibitem{Bhandari:Crystals2021}
U. Bhandari, C. Zhang, C. Zeng, S. Guo, A. Adhikari, S. Yang, Crystals 11 (2021) 46.

\bibitem{Lin:JALCOM2015}
C.-M. Lin, C. -C. Juan, C. -H. Chang, C. -W. Tsai, J. -W. Yeh, J. Alloys Compd. 624 (2015) 100.

\end{thebibliography}
\end{document}